\documentclass[preprint,12pt]{elsarticle} 
\usepackage{amsmath,amssymb,graphicx,xcolor,stix2,bm,listings,braket}
\usepackage[colorlinks=true, allcolors=blue]{hyperref}
\usepackage[htt]{hyphenat}
\usepackage{ulem}   
\biboptions{square,numbers,comma,sort&compress}

\newcommand{\bn}[1]{\mathbf{#1}}    
\newcommand{\bi}[1]{\bm{#1}}        
\renewcommand{\rm}[1]{\mathrm{#1}}    
\newcommand{\dd}{\mathrm{d}}    
\newcommand{\ii}{\mathrm{i}}    
\newcommand{\ee}{\mathrm{e}}    
\newcommand{\abs}[1]{\lvert #1 \rvert}  

\journal{Computer Physics Communications}

\begin{document}

\begin{frontmatter}

\title{PyStructureFactor:\texorpdfstring{\\}{ }A Python code for the molecular structure factor in tunneling ionization rates}

\author[a]{Shanshan Song\fnref{contrib}}
\author[a,b]{Mingyu Zhu\fnref{contrib}}
\author[a,c,d]{Hongcheng Ni\corref{cor1}}
\author[a,c,d,e]{Jian Wu\corref{cor2}}

\address[a]{State Key Laboratory of Precision Spectroscopy, East China Normal University, Shanghai 200241, China}
\address[b]{School of Physics and Electronic Science, East China Normal University, Shanghai 200241, China}
\address[c]{NYU-ECNU Joint Institute of Physics, New York University at Shanghai, Shanghai 200062, China}
\address[d]{Collaborative Innovation Center of Extreme Optics, Shanxi University, Taiyuan, Shanxi 030006, China}
\address[e]{CAS Center for Excellence in Ultra-intense Laser Science, Shanghai 201800, China}

\cortext[cor1]{Corresponding author.\\\textit{E-mail address:} hcni@lps.ecnu.edu.cn}
\cortext[cor2]{Corresponding author.\\\textit{E-mail address:} jwu@phy.ecnu.edu.cn}
\fntext[contrib]{These authors contributed equally.}

\begin{abstract}
Tunneling ionization is at the core of strong-field and attosecond science. In this paper, we present \texttt{PyStructureFactor} --- a general \texttt{Python} code towards the calculation of the structure factor in the tunneling ionization rate of common molecules under intense laser fields. The numerical implementation is based on the well-developed weak-field asymptotic theory in the integral representation. The information of the electronic structure of the molecules is obtained via the \texttt{PySCF} quantum chemistry package. \texttt{PyStructureFactor} is a general computational framework that can be utilized to compute the molecular structure factor of various types of molecules, including polar and nonpolar diatomic molecules, degenerate molecules, and open-shell molecules. Examples are given that are benchmarked against known results with good agreements. The present \texttt{PyStructureFactor} is implemented in an efficient manner and is easily applicable towards larger molecules.
\end{abstract}


\end{frontmatter}

{\bf \noindent PROGRAM SUMMARY} \\
\begin{small}
\noindent
{\em Program Title:} PyStructureFactor \\
{\em CPC Library link to program files:} (to be added by Technical Editor) \\
{\em Developer's repository link:} \url{https://github.com/TheStarAlight/PyStructureFactor} \\
{\em Code Ocean capsule:} (to be added by Technical Editor) \\
{\em Licensing provisions:} Apache-2.0 \\
{\em Programming language:} Python 3                                  \\
{\em Supplementary material:}                                 \\
{\em Nature of problem:} The structure factor of a molecule in intense laser fields determines its orientation-dependent tunneling ionization rate, which is crucial for the studies of ultrafast molecular dynamics and its control. However, the strong-field community lacks an open-source code to calculate the molecular structure factor, and can only resort to known results of a few molecules. \\
{\em Solution method:} We developed the \texttt{PyStructureFactor} program with the structure factor of arbitrary molecules calculated using the weak-field asymptotic theory in the integral representation. The underlying electronic structure necessary for the calculation is obtained using the \texttt{PySCF} quantum chemistry package. \\
{\em Restrictions on the Accuracy:} The accuracy of the molecular structure factor calculated by \texttt{PyStructureFactor} is restricted by the level of precision of the electronic structure information extracted from the \texttt{PySCF} package. \\
{\em Running time:} The running time depends on the size of the molecule, the basis set of the calculation, the level of precision of the electronic structure calculations, and other parameters passed to the program. The example in Fig.~\ref{fig:code_example} took 1.2 seconds to finish on an AMD Ryzen 9 7950X CPU on the WSL Ubuntu 22.04 LTS.
\end{small}

\section{Introduction}
\label{sec:intro}

Advances in the laser technology have made intense laser fields widely accessible in typical table-top lab settings. When matter interacts with intense laser fields, a plethora of novel strong-field phenomena emerge that are of fundamental interest, such as high-order harmonic generation (HHG) \cite{krause_high_1992,corkum_plasma_1993,popmintchev_the_2010}, above-threshold ionization \cite{agostini_free_1979,becker_above_2002,milosevic_above_2006}, nonsequential double ionization \cite{walker_precision_1994}, and laser-induced electron diffraction \cite{blaga_imaging_2012,wolter_ultrafast_2016}. HHG is the key to produce extreme ultraviolet light pulses, which has facilitated the improvement of time resolution of ultraprecise measurements down to the attosecond level \cite{krausz_attosecond_2009,dahlstroem_introduction_2012,pazourek_attosecond_2015,kheifets_wigner_2023}. Common to many of these strong-field phenomena is their first step, known as tunneling ionization \cite{keldysh_ionization_1965,chin_from_2004}, which is thus a cornerstone of strong-field and attosecond science. It has attracted widespread interest not only in physics but also in chemistry and related fields \cite{smirnova_attosecond_2009,sukiasyan_exchange_2010,xie_attosecond-recollision-controlled_2012,spanner_strong-field_2012,doblhoff-dier_classical_2013,doblhoff-dier_theoretical_2016,patchkovskii_full-dimensional_2017}. Given the recent development in optochemistry \cite{li_light_2022,ma_transient_2021}, a complementary emerging field to photochemistry, the importance of the tunneling ionization of molecules has been elevated to new heights.

In contrast to photoionization where the photonic feature of light plays a dominant role, tunneling ionization is better represented by the highly nonlinear ionization induced by intense optical fields in the time domain $F(t)$, with a tunneling ionization rate \cite{ammosov_tunnel_1986,delone_tunneling_1998}
\begin{equation}
    W[F(t)] \sim \ee^{-2\kappa^3/3F(t)}
    \label{eq:FieldFactor}
\end{equation}
depending exponentially on the field strength $F(t)$, where $\kappa=\sqrt{-2E_0}$ with $E_0$ the orbital energy of the target. For typical intense infrared laser fields, the photon energy is substantially lower than the ionization potential of the bound electron. Moreover, the timescale of the field oscillation is much longer than the intrinsic attosecond timescale of the bound electron. Therefore, the strong-field ionization process can be approximated by tunneling in a static electric field represented by the instantaneous laser electric field $F(t)$ \cite{tolstikhin_adiabatic_2010, tolstikhin_adiabatic_2012}.

In addition to the field factor [Eq.~(\ref{eq:FieldFactor})], the tunneling ionization rate of molecules depends on a factor related to the electronic structure of the ionizing orbital, known as the molecular structure factor. It determines the orientation-dependent rate of tunneling ionization that is crucial to the analysis of subsequent strong-field molecular dynamics and photoelectron spectrum \cite{pavicic_direct_2007}. It is the aim of the current work to present the numerical solution to the molecular structure factor.

The molecular structure factor can be obtained either by the molecular Ammosov--Delone--Krainov theory (MO-ADK) \cite{tong_theory_2002, pavicic_direct_2007, holmegaard_photoelectron_2010, johansen_alignment-dependent_2016, zhao_accurate_2017} or via the weak-field asymptotic theory (WFAT) \cite{tolstikhin_theory_2011, madsen_application_2012, madsen_structure_2013, trinh_first-order_2013, madsen_application_2014, tolstikhin_weak-field_2014, tolstikhina_application_2014, saito_structure_2015, trinh_weak-field_2015, dnestryan_integral-equation_2016, svensmark_theory_2016, trinh_first-order_2016, madsen_structure_2017, dnestryan_structure_2018, samygin_weak-field_2018, matsui_weak-field_2021}. WFAT is applicable in laser fields below the critical field strength that leads to over-barrier ionization. Based on the parabolic adiabatic expansion approach \cite{batishchev_atomic_2010}, WFAT generalizes isotropic atomic potentials to arbitrary molecular potentials and naturally accounts for the effect of molecular permanent dipole moment (if any), and has been adopted as the theoretical method in some experimental research on molecular strong-field ionization \cite{ohmura_molecular_2014, kraus_observation_2015, walt_role_2015, endo_imaging_2016, fujise_helicity-dependent_2022}. In this work, we employ WFAT towards a numerical implementation of the computation of the molecular structure factor. Under its framework, the obtained tunneling ionization rate, in its leading order, factorizes into the field factor [Eq.~(\ref{eq:FieldFactor})] and the absolute square of the structure factor, the latter of which is an inherent property of the ionizing molecular orbital independent of the external field \cite{trinh_first-order_2013}.

In an essence of the perturbation theory, the structure factor of WFAT is found through the behavior of the wave function in the asymptotic region in early stages, known as the tail representation \cite{madsen_structure_2013, trinh_weak-field_2015, madsen_application_2014}. Hence, its reliable evaluation requires an accurate description of the asymptotic tail of the ionizing orbital. To extract the orbital wave function of polyatomic molecules by the Hartree-Fock (HF) method or density functional theory (DFT) with quantum chemistry software packages \cite{madsen_application_2014, saito_structure_2015}, one normally resorts to the basis-based approaches. However, compared to the correct exponential behavior obtained by the grid-based approach, the standard Gaussian-type basis functions facilitating multicenter integral calculations decay too rapidly in the asymptotic region and thus could result in erratic oscillations. To reproduce the correct local property of the asymptotic tail of the orbital, large basis sets with quadruple or pentuple-zeta quality and variationally optimized exponents are required \cite{trinh_first-order_2013}. This approach significantly increases the computational cost and become infeasible in many cases. The efficiency of such numerical computation needs to be improved urgently.

As a remedy, an alternative integral approach of WFAT was developed with the spirit of scattering theory to overcome the above issue \cite{madsen_structure_2017, dnestryan_structure_2018, samygin_weak-field_2018}. In this approach, the structure factor is expressed as an integral form involving the ionizing orbital and a known analytical function. The integral accumulates at the localized regions where the orbital has large amplitudes, and it is insensitive to the representation of the asymptotic tail far from the nuclei. Hence, the integral approach can be implemented using standard quantum chemistry software packages with Gaussian-type orbitals (GTOs) for arbitrary molecules.

In this work, an open-source program implementing WFAT in the integral representation is developed that can easily and quickly calculate the molecular structure factor of common molecules. In our implementation, named the \texttt{PyStructureFactor}, the ionizing molecular orbital is expanded as a linear combination of GTOs at the HF or complete-active-space self-consistent field (CASSCF) level of theory calculated by the \texttt{PySCF} package \cite{sun_pyscf_2018, sun_recent_2020}. The integrals are evaluated numerically using a three-dimensional grid-based method, which exhibits much better numerical stability and is readily applicable to large molecules, compared to the standard analytical integration procedure. Our present implementation of \texttt{PyStructureFactor} features flexible control of accuracy and efficient computation towards the calculation of the molecular structure factor.

This work is organized as follows. In Sec.~\ref{sec:theory}, the definition and derivation of the tunneling ionization rate, the molecular structure factor, and other parameters within the WFAT framework are given. In Sec.~\ref{sec:numerical}, the details of numerical implementation are presented. In Sec.~\ref{sec:usage}, we detail the usage of the \texttt{PyStructureFactor} program to calculate the molecular structure factor. In Sec.~\ref{sec:example}, we illustrate the application of the \texttt{PyStructureFactor} program with examples of various types of common molecules. Conclusions are given in Sec.~\ref{sec:concl}. Atomic units are used throughout unless stated otherwise.

\section{Theoretical formulation}
\label{sec:theory}

Tunneling ionization of molecules can be usually modeled within the Born-Oppenheimer and the single-active-electron approximation. Under such a framework, the wave function of the ionizing orbital $\psi_0$ is the solution to an effective one-electron Schr\"{o}dinger equation
\begin{equation}
    \left[-\frac12\nabla^2 + V(\bn{r})\right] \psi_0(\bn{r}) = E_0 \psi_0(\bn{r}),
\end{equation}
where the HF potential $V(\bn{r})$ describes the interaction between the ionizing electron and the parent ion.

\begin{figure}[tb]
    \centering
    \includegraphics[width=0.5\columnwidth]{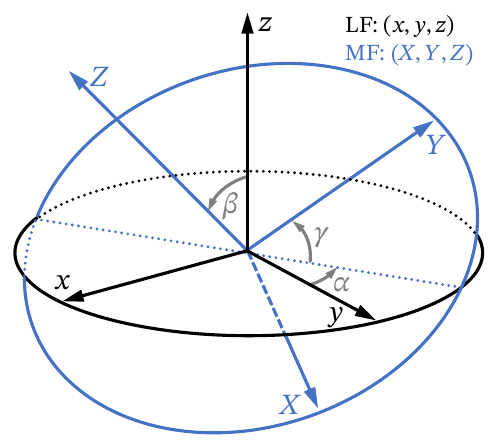}
    \caption{
        Illustration of the rotation from the LF to the MF using Euler angles of $z$-$y'$-$z''$ convention. The rotation consists of three steps: (1) The coordinate system $(x,y,z)$ rotates about the $z$ axis by angle $\alpha$, then (2) rotates about the new $y$ axis by angle $\beta$, finally (3) rotates about the rotated $z$ axis by $\gamma$, and becomes $(X,Y,Z)$. The rotations mentioned above are all counter-clockwise.
        }
    \label{fig:EulerAngle}
\end{figure}

Within WFAT, tunneling ionization is investigated in parabolic coordinates. We assume the electric field vector $\bn{F}=F\hat{\bn{z}}$, and define the orientation of the molecule using a set of Euler angles $(\alpha,\beta,\gamma)$ (within the $z$-$y'$-$z''$ convention \cite{madsen_structure_2017}, which is shown schematically in Fig.~\ref{fig:EulerAngle}) representing a rotation $\hat{\bn{R}}$ from the laboratory frame (LF) to the molecular frame (MF). The ionization rate does not depend on $\alpha$ and so we set $\alpha=0$ throughout. The total ionization rate is split into different parabolic channels:
\begin{equation}
    \Gamma(\beta, \gamma)=\sum_\nu \Gamma_\nu(\beta, \gamma),
\end{equation}
where $\Gamma_\nu(\beta, \gamma)$ are partial rates of parabolic quantum number indices
\begin{equation}
    \nu=(n_\xi,m), \quad n_\xi=0,1,2,\cdots, \quad m=0,\pm 1,\pm 2,\cdots
\end{equation}
In the leading-order approximation, the asymptotic expansion of the partial rates for $F\rightarrow 0$ gives
\begin{equation}
    \Gamma_{\nu}(\beta, \gamma) = \abs{G_\nu(\beta, \gamma)}^2 W_{\nu}(F),
\end{equation}
which can be separated into two factors: The structural part $\abs{G_\nu(\beta, \gamma)}^2$ and the field part $W_{\nu}(F)$.
The field factor, $W_{\nu}(F)$, has the form \cite{ammosov_tunnel_1986,delone_tunneling_1998}
\begin{equation}
    W_{\nu}(F) = \frac{\kappa}{2} \left(\frac{4\kappa^2}{F}\right)^{2Z/\kappa-2n_\xi-\abs{m}-1} \ee^{-{2\kappa^3}/{3F}}.
\end{equation}
The structure factor, namely $G_\nu(\beta, \gamma)$, in the integral representation of WFAT, is given as an integral:
\begin{equation}
    G_\nu (\beta,\gamma) = \ee^{-\kappa\mu_z} \int \Omega_\nu^* \left(\hat{\bn{R}}^{-1} \bn{r}\right) V_{\rm{c}}(\bn{r}) \psi_0(\bn{r}) \dd \bn{r},
    \label{eq:StructFactorExp}
\end{equation}
where
\begin{equation}
    \bi{\mu} = \int \psi_0^*(\bn{r}) \bn{r} \psi_0(\bn{r}) \dd \bn{r}
\end{equation}
denotes the orbital dipole moment in the LF with $\mu_z$ its component along the field direction;
$\Omega_\nu$ is a reference function which can be expanded into spherical harmonics:
\begin{equation}
    \Omega_\nu(\bn{r}) = \sum_{l=|m|}^{\infty} \Omega^\nu_{lm}(\bn{r}),
    \label{eq:OmegaExpansion1}
\end{equation}
where
\begin{equation}
    \Omega^\nu_{lm}(\bn{r}) = \sum_{l=|m|}^{\infty} R_l^\nu(r) Y_{lm}(\theta, \phi);
    \label{eq:OmegaExpansion2}
\end{equation}
$V_{\rm{c}}(\bn{r}) = V(\bn{r}) + Z/r$ is the core potential without the Coulomb tail (i.e., $V_{\rm{c}} \rightarrow 0$ as $r\rightarrow \infty$) with $Z$ the asymptotic charge of the parent molecular ion.
The radial part $R_l^\nu$ of the reference function $\Omega_{lm}^\nu$ in Eq.~(\ref{eq:OmegaExpansion2}) reads
\begin{equation}
    R_l^\nu(r)=\omega_l^\nu \ (\kappa r)^l \ \ee^{-\kappa r} \ \rm{M}(l+1-Z/\kappa, 2l+2, 2 \kappa r),
\end{equation}
where $\rm{M}(a, b, x)$ is the confluent hyper-geometric function \cite{olver_nist_2010} and $\omega_l^\nu$ is the normalization coefficient:
\begin{align}
    \omega_l^\nu = & \      (-1)^{l+(\abs{m}-m)/2+1}\ 2^{l+3/2}\ \kappa^{Z/\kappa-(\abs{m}+1)/2-n_\xi} \nonumber \\
                   & \times \sqrt{(2l+1)(l+m)!(l-m)!(\abs{m}+n_\xi)!n_\xi!}\ \frac{l!}{(2l+1)!} \nonumber \\
                   & \times \!\!\!\!\!\! \sum_{k=0}^{\min{(n_\xi,l-\abs{m})}} \!\!\!\!\!\!\!\!\!\! \frac{\Gamma(l+1-Z/\kappa+n_\xi-k)}{k!(l-k)!(\abs{m}+k)!(l-\abs{m}-k)!(n_\xi-k)!}.
\end{align}

Under the framework of the HF method, the HF potential $V(\bn{r})$ consists of three parts, namely the nuclear Coulomb potential ($V_{\rm{nuc}}$), the direct ($V_{\rm{d}}$) and exchange ($V_{\rm{ex}}$) parts of inter-electron interactions:
\begin{equation}
    V(\bn{r}) = V_{\rm{nuc}}(\bn{r}) + V_{\rm{d}}(\bn{r}) + V_{\rm{ex}}(\bn{r}),
\end{equation}
where
\begin{align}
    V_{\rm{nuc}}(\bn{r}) &= -\sum_{A=1}^{N_{\rm{atm}}} \frac{Z_A}{\left|\bn{r}-\bn{R}_A\right|}, \label{eq:Vnuc} \\
    V_{\rm{d}}(\bn{r}) &= \sum_{i=1}^N \int \frac{\psi_i^*(\bn{r}') \psi_i(\bn{r}')}{\abs{\bn{r}-\bn{r}'}} \dd \bn{r}', \label{eq:Vd} \\
    V_{\rm{ex}}(\bn{r}) \psi_0(\bn{r}) &= -\sum_{i=1}^N \psi_i(\bn{r}) \int \frac{\psi_i^*(\bn{r}') \psi_0(\bn{r}')}{\abs{\bn{r}-\bn{r}'}} \braket{\sigma_i | \sigma_0} \dd \bn{r}', \label{eq:VexPsi0}
\end{align}
where $N$ is the number of electrons, $N_{\rm{atm}}$ is the number of nuclei, $\psi_i(\bn{r})$ and $\sigma_i$ denote the molecular orbital and the spin state of the electron of index $i$ ($\braket{\sigma_i | \sigma_j}=1$ if electrons $i$ and $j$ have the same spin, and $\braket{\sigma_i | \sigma_j}=0$ otherwise), while $Z_A$ and $\bn{R}_A$ are the nuclear charge and position of atom of index $A$.

\section{Numerical implementation}
\label{sec:numerical}

For the practical numerical implementation of the computation of the molecular structure factor within the framework of WFAT, we first refer to the evaluation of the integral in Eq.~(\ref{eq:StructFactorExp}). Directly evaluating the integrals for each molecular orientation $(\beta,\gamma)$ is inefficient. Nevertheless, we may utilize the spherical harmonic expansion of the reference function $\Omega_\nu$, and express the rotated spherical harmonic functions as a linear combination of spherical harmonic functions using the Wigner-$d$ matrix. Substituting Eqs.~(\ref{eq:OmegaExpansion1}) and (\ref{eq:OmegaExpansion2}) into Eq.~(\ref{eq:StructFactorExp}) yields \cite{madsen_structure_2017}
\begin{equation}
    G_\nu(\beta,\gamma) = \ee^{-\kappa \mu_z} \sum_{l=|m|}^{\infty} \sum_{m'=-l}^l I_{l m'}^\nu  d_{m m'}^l(\beta) e^{-\ii m' \gamma},
    \label{eq:StructFactorExpInt}
\end{equation}
where $d_{m m'}^l(\beta)$ is the Wigner-$d$ matrix, and the integral $I_{l m'}^\nu$ is expressed as
\begin{equation}
    I_{l m'}^\nu = \int \Omega_{l m'}^{\nu *}(\bn{r}) V_{\rm{c}}(\bn{r}) \psi_0(\bn{r}) \dd \bn{r}.
    \label{eq:IntegralExpr}
\end{equation}
In this way the computational overhead is significantly reduced --- with the values of the integrals $I_{l m'}^\nu$, we may obtain the structure factor of arbitrary molecular orientation $G_\nu(\beta,\gamma)$ with little computational cost.

The next topic is concerned with the calculation of the HF potential operator acting on the ionizing orbital, i.e., $V(\bn{r}) \psi_0(\bn{r})$. In the practical application, the molecular orbitals are usually obtained by the self-consistent-field (SCF) approach, which is commonly implemented in standard quantum chemistry packages. The molecular orbitals are usually expanded into linear combinations of Gaussian-type orbitals (GTOs):
\begin{equation}
    \psi_i(\bn{r})=\sum_\alpha C_{i \alpha} \chi_\alpha(\bn{r}-\bn{R}_\alpha),
\end{equation}
where $\chi_\alpha(\bn{r}-\bn{R}_\alpha)$ denotes a GTO of index $\alpha$ centering at $\bn{R}_\alpha$, and the coefficients $C_{i \alpha}$ are obtained through the SCF approach.
In this way we obtain $V_{\rm{d}}(\bn{r})$ and $V_{\rm{ex}}(\bn{r})\psi_0(\bn{r})$ in Eqs.~(\ref{eq:Vd}) and (\ref{eq:VexPsi0}) expressed in coefficients $C_{i \alpha}$ for practical calculations. Defining the integral
\begin{equation}
    J_{\alpha\beta}(\bn{r}) = \int \frac{\chi_\alpha(\bn{r}'-\bn{R}_\alpha) \chi_\beta(\bn{r}'-\bn{R}_\beta)}{\abs{\bn{r}-\bn{r}'}} \dd \bn{r}'
\end{equation}
which can be calculated analytically for GTOs,
Eqs.~(\ref{eq:Vd}) and (\ref{eq:VexPsi0}) can be written as
\begin{align}
     V_{\rm{d}}(\bn{r}) &= \sum_{i=1}^{N} \sum_{\alpha,\beta} C_{i\alpha} C_{i\beta} J_{\alpha\beta}(\bn{r}), \\
     V_{\rm{ex}}(\bn{r})\psi_0(\bn{r}) &= - \sum_{i=1}^{N} \braket{\sigma_i | \sigma_0} \sum_{\alpha,\beta,\gamma} C_{i\alpha} C_{i\beta} C_{0\gamma} J_{\beta\gamma}(\bn{r}) \chi_\alpha(\bn{r}-\bn{R}_\alpha).
\end{align}

The last topic is related to the convergence of the algorithm. The integral in Eq.~(\ref{eq:IntegralExpr}) is numerically evaluated on a grid using the Becke fuzzy cell integration scheme \cite{becke_multicenter_1988} implemented in the \texttt{PySCF} package, which uses higher density of grid points near the nuclei to accelerate convergence of the calculation. Apart from adopting the Becke grids in the numerical integration, we also choose the origin such that the dipole of the parent ion
\begin{equation}
    \bn{D} = \left[\sum_{A=1}^{N_{\rm{atm}}} Z_A \bn{R}_A - \sum_{i=1}^N \int \psi_i^*(\bn{r}) \bn{r} \psi_i(\bn{r}) \dd \bn{r} \right] - \bi{\mu}
    \label{eq:ParentIonDipole}
\end{equation}
vanishes, which reduces the grid size required for the integral Eq.~(\ref{eq:IntegralExpr}) to converge.

\section{Program usage}
\label{sec:usage}

\begin{table}[tb]
    \small
    \centering
    \begin{tabular}{l p{9cm} l}
        \hline \hline
        Parameter                   & Description                                                                                       & Default\\
        \hline
        \texttt{atom}               & Information of the molecular structure. Refer to Fig.~\ref{fig:code_example} for an example.      & ---\\
        \texttt{unit}               & The unit used for the position coordinates of the atoms in the molecule.
                                      \texttt{'B'} or \texttt{'AU'} indicates Bohr and otherwise Angstr\"om.                            & \texttt{'Angstr\"om'}\\
        \texttt{basis}              & Basis set used for calculation.
                                      Refer to \href{https://pyscf.org/pyscf_api_docs/pyscf.gto.basis.html}{\texttt{pyscf.gto.basis}}
                                      for more information.
                                      An accurate basis set is suggested for optimal convergence.                                       & \texttt{'sto-3g'}\\
        \texttt{charge}             & Charge of the molecule or molecular ion.                                                          & \texttt{0} \\
        \texttt{spin}               & Number of spin (i.e., $2S$) of the molecule.                                                      & \texttt{0} \\
        \hline \hline
    \end{tabular}
    \caption{Essential input parameters of the \texttt{pyscf.M} method in the \texttt{PyStructureFactor} program.}
    \label{tab:pyscfm_param}
\end{table}

\begin{table}[tb]
    \small
    \centering
    \begin{tabular}{l p{9cm} l}
        \hline \hline
        Parameter                   & Description                                                                                               & Default           \\
        \hline
        \texttt{mol}                & The \texttt{PySCF} molecule object. Initialized by invoking \texttt{pyscf.M} or
                                      \href{https://pyscf.org/pyscf_api_docs/pyscf.gto.html#module-pyscf.gto.mole}{\texttt{pyscf.gto.M}}.
                                                                                                                                                & ---               \\
        \texttt{orbital\_index}     & Index of the ionizing orbital relative to HOMO, e.g., $\rm{HOMO} \rightarrow 0$,
                                      $\rm{LUMO} \rightarrow +1$, $\rm{HOMO-1} \rightarrow -1$, …
                                                                                                                                                & \texttt{0}        \\
        \texttt{channel}            & Parabolic channel $\nu=(n_\xi, m)$.
                                      The program would calculate the structure factor of channel $(n_\xi, m)$, i.e., $G_{n_\xi,m}$.
                                                                                                                                                & \texttt{(0,0)}    \\
        \texttt{lmax}               & The cut-off limit of the angular quantum number (larger $l$ would be neglected)
                                      used in the summation in Eq.~(\ref{eq:StructFactorExpInt}).
                                                                                                                                                & \texttt{10}       \\
        \texttt{hf\_method}         & Indicates whether \texttt{'RHF'} or \texttt{'UHF'} should be used in the molecular HF calculation.
                                      Note: \texttt{'UHF'} must be used for open-shell molecules.
                                                                                                                                                & \texttt{'RHF'}    \\
        \texttt{casscf\_conf}       & Configuration of CASSCF calculation consisting of \texttt{(n\_active\_orb, n\_active\_elec)}.
                                      Specifying \texttt{None} (by default) indicates employing the primitive HF method instead of the more accurate CASSCF method.
                                                                                                                                                & \texttt{None}\\
        \texttt{atom\_grid\_level}  & Level of fineness of the grid used in integration
                                      (see also \href{https://pyscf.org/pyscf_api_docs/pyscf.dft.html#module-pyscf.dft.gen_grid}{\texttt{pyscf.dft.Grid}}),
                                      which controls the number of radial and angular grids around each atom in the evaluation of the integration,
                                      ranging from 0 to 9.
                                                                                                                                                & \texttt{3}        \\
        \texttt{orient\_grid\_size} & Size of the output $(\beta,\gamma)$ grid, which defines the orientation of the molecule with respect to the polarization direction of the laser field.
                                      The grid is uniform, with $\beta$ ranging from $0$ to $\pi$ and $\gamma$ ranging from $0$ to $2\pi$.
                                      Setting the $\gamma$ grid count to $1$ indicates that $\gamma$ would be zero throughout the calculation.
                                                                                                                                                & \texttt{(90,1)}   \\
        \texttt{move\_dip\_zero}    & Indicates whether to shift the molecular coordinates such that the dipole moment of the parent ion [Eq.~(\ref{eq:ParentIonDipole})] vanishes
                                      (i.e., $\bn{D}=\bn{0}$).
                                                                                                                                                & \texttt{True}     \\
        \texttt{rmax}               & Indicates the cut-off limit of the radial grid points,
                                      points of radii larger than \texttt{rmax} would be neglected in the calculation.
                                                                                                                                                & \texttt{40}       \\
        \hline
        Returns                     & A \texttt{NumPy} array containing the structure factor $G_{n_\xi,m}$ of the given channel $(n_\xi,m)$
                                      on the $(\beta,\gamma)$ orientation grid, whose shape is given by \texttt{orient\_grid\_size}. \\
        \hline \hline
    \end{tabular}
    \caption{Input parameters and return value of the \texttt{PyStructureFactor.get\_structure\_factor} method.}
    \label{tab:getsf_param}
\end{table}

The usage of the program is simple and straightforward.
Firstly, import necessary packages, including \texttt{PyStructureFactor}, \texttt{PySCF} and \texttt{NumPy} in Python.
Next, initialize the molecule object by calling the method \texttt{pyscf.M}, and provide necessary information of the molecule.
Finally, get the structure factor of the molecule by invoking the method \texttt{PyStructureFactor.get\_structure\_factor}, whose return value is a \texttt{NumPy} array containing the structure factor at different molecular orientations $(\beta,\gamma)$.

To better illustrate the usage of the program, we present a description of the input parameters of the two main methods, including \texttt{pyscf.M}, which defines the molecular and orbital information, and \texttt{PyStructureFactor.get\_structure\_factor}, which outputs the corresponding structure factor, in Tables~\ref{tab:pyscfm_param} and \ref{tab:getsf_param}, respectively.
Shown in Fig.~\ref{fig:code_example} is a minimal example of the program which calculates the molecular structure factor $G_{00}(\beta)$ corresponding to the highest occupied molecular orbital (HOMO) of the hydrogen molecule. For the diatomic molecules, the $\gamma$ Euler angle does not come into play, and $\beta$ denotes the angle between the electric field vector and the molecular axis in the LF.
More examples are available in Sec.~\ref{sec:example}.

\begin{figure}[tb]
    \begin{lstlisting}[language=Python, basicstyle=\footnotesize\ttfamily, breaklines=true, keywordstyle=\color{blue}, stringstyle=\color{purple}, commentstyle=\color{gray}, frame=tb, framesep=1em]
from PyStructureFactor import get_structure_factor
import numpy as np
import pyscf
n_beta  = 90
n_gamma = 1
molH2 = pyscf.M(atom="H 0,0,0.37; H 0,0,-0.37", basis="pc-1", spin=0)
beta_grid = np.linspace(0, np.pi, n_beta)
G_grid = get_structure_factor(
            mol = molH2, orbital_index = 0, channel = (0,0),
            lmax = 10, hf_method = "RHF",
            atom_grid_level = 3,
            orient_grid_size = (n_beta, n_gamma))
    \end{lstlisting}
    \caption{A minimal example of the program calculating the structure factor $G_{00}(\beta)$ of the HOMO of the hydrogen molecule ($\rm{H}_2$).}
    \label{fig:code_example}
\end{figure}

\begin{figure}[tb]
    \centering
    \includegraphics[width=\columnwidth]{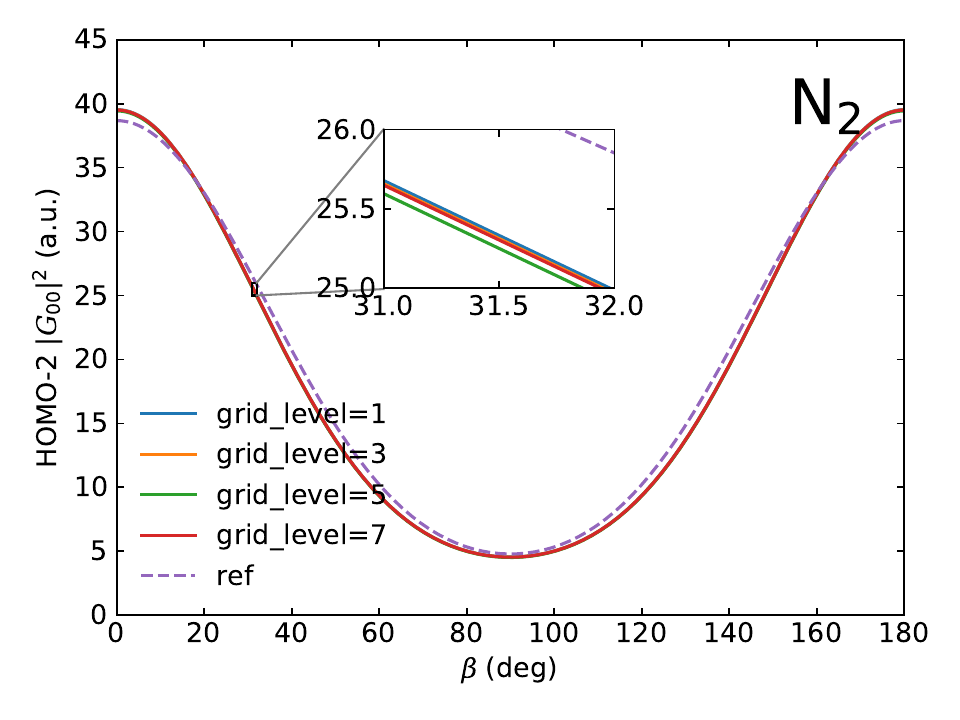}
    \caption{Squared structure factor $\abs{G_{00}}^2$ of the HOMO-2 of the nitrogen molecule ($\rm{N}_2$). The reference data are extracted from Ref.~\cite{saito_structure_2015}.}
    \label{fig:convtest_N2}
\end{figure}

\begin{figure}[tb]
    \centering
    \includegraphics[width=\columnwidth]{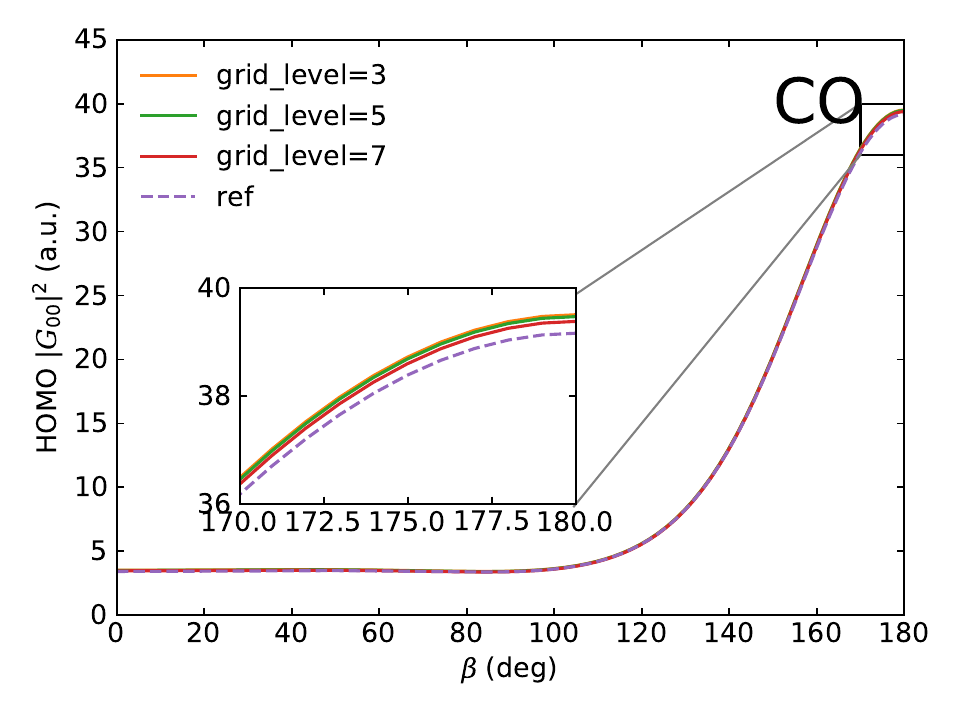}
    \caption{Squared structure factor $\abs{G_{00}}^2$ of the HOMO of the carbon monoxide molecule ($\rm{CO}$). The reference data are extracted from Ref.~\cite{saito_structure_2015}.}
    \label{fig:convtest_CO}
\end{figure}

Now we present convergence and reliability tests of the program in Figs.~\ref{fig:convtest_N2} and \ref{fig:convtest_CO} with N$_2$ and CO molecules, respectively. To this end, we obtain their structure factors as a function of $\beta$ and compare the results to reference data extracted from the literature \cite{saito_structure_2015} which is obtained using the tail representation of WFAT. In both figures, the pc-4 basis is adopted in the calculations, and the cut-off limit of the index $l$ in the summation of Eq.~(\ref{eq:StructFactorExpInt}) , i.e., $l_{\rm{max}}$, is set to the default value of 10. The ``grid\_level'' labels denote the input parameter \texttt{``atom\_grid\_level''} of the program. Here, grid levels 1, 3, 5, and 7 correspond to radial and angular sizes (40, 194), (75, 302), (105, 770), and (135, 1202) for each spherical grid around the nuclei, respectively. It is clear from the figures that for \texttt{``atom\_grid\_level''} larger than 3, the obtained results are hardly distinguishable from each other even after enlargement, which reveals a stable and fast convergence of the present program. In addition, the converged result is close to the reference data. This illustrates that the calculation of the molecular structure factor is able to achieve accurate results under default parameters. Minor discrepancy from the reference data results from the difference in the tail and integral presentations of WFAT and the choice of basis.

\section{Illustrative examples}
\label{sec:example}

\begin{figure}
    \centering
    \includegraphics[width=\columnwidth]{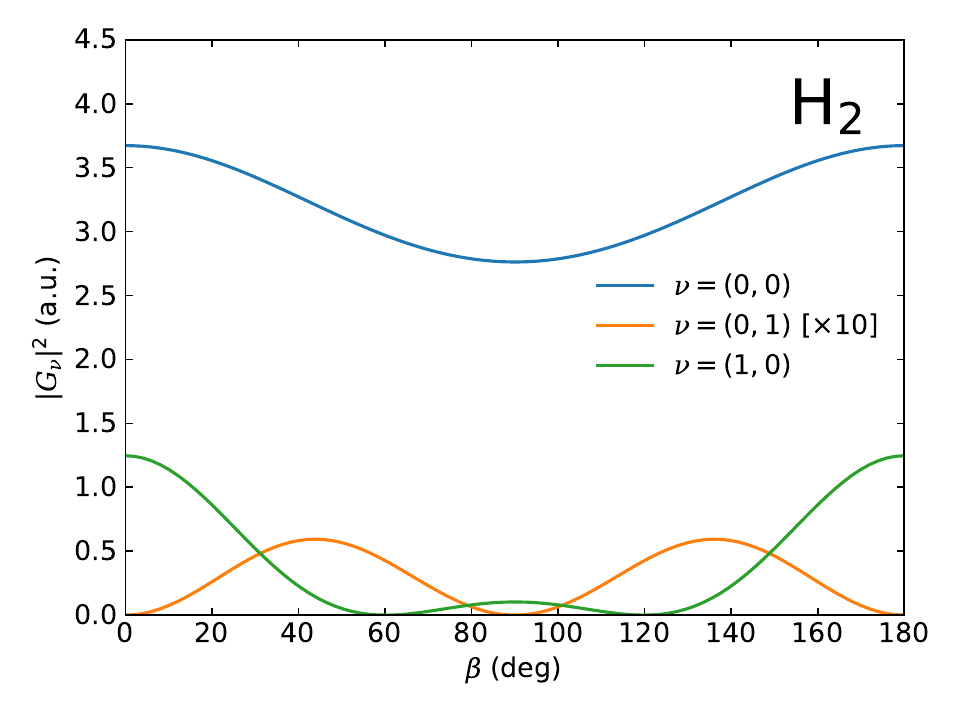}
    \caption{Squared structure factor $\abs{G_\nu}^2$ of the HOMO of the hydrogen molecule ($\rm{H}_2$). The molecular axis is aligned along the $z$-axis in the MF and the internuclear distance is 0.74 \r{A}. The basis used for calculation is pc-4.}
    \label{fig:example_H2}
\end{figure}

\begin{figure}
    \centering
    \includegraphics[width=\columnwidth]{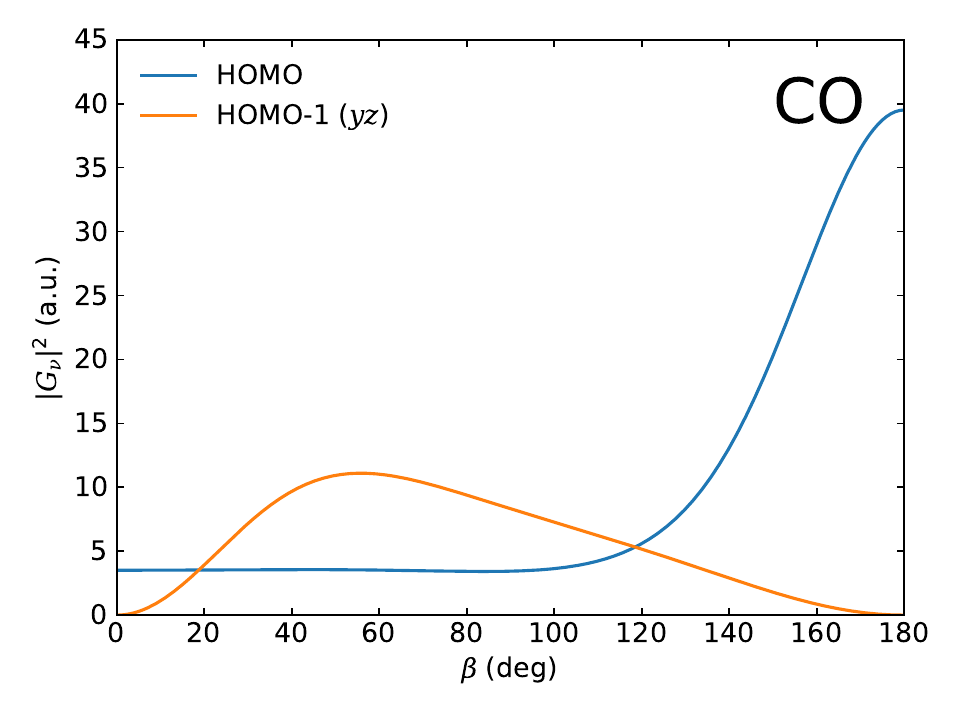}
    \caption{Squared structure factor $\abs{G_{00}}^2$ of the HOMO and HOMO-1 of the carbon monoxide molecule ($\rm{CO}$, with the nodal plane on the $y$-$z$ plane). The molecular axis is aligned along the $z$-axis in the MF with the $+z$ axis pointing from the C to the O nucleus, and the internuclear distance is 1.13 \r{A}. The basis used for calculation is pc-4.}
    \label{fig:example_CO}
\end{figure}

\begin{figure}
    \centering
    \includegraphics[width=\columnwidth]{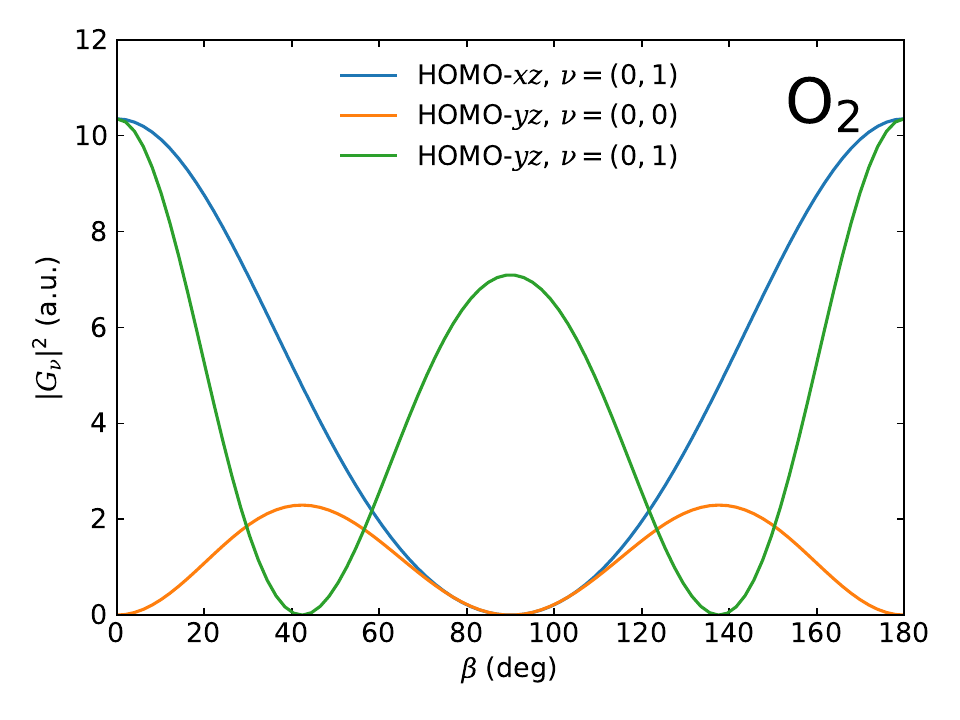}
    \caption{Squared structure factor $\abs{G_\nu}^2$ of the two degenerate HOMOs of the oxygen molecule ($\rm{O}_2$, denoted by their nodal planes $xz$ and $yz$). The molecular axis is aligned along the $z$-axis in the MF and the internuclear distance is 1.21 \r{A}. The basis used for calculation is pc-4. The oxygen molecule at the ground state has two unpaired electrons and thus has a total spin $S=1$, which requires parameters \texttt{spin=2} and \texttt{hf\_method="UHF"} to be passed to the program.}
    \label{fig:example_O2}
\end{figure}

\begin{figure}
    \centering
    \includegraphics[width=\columnwidth]{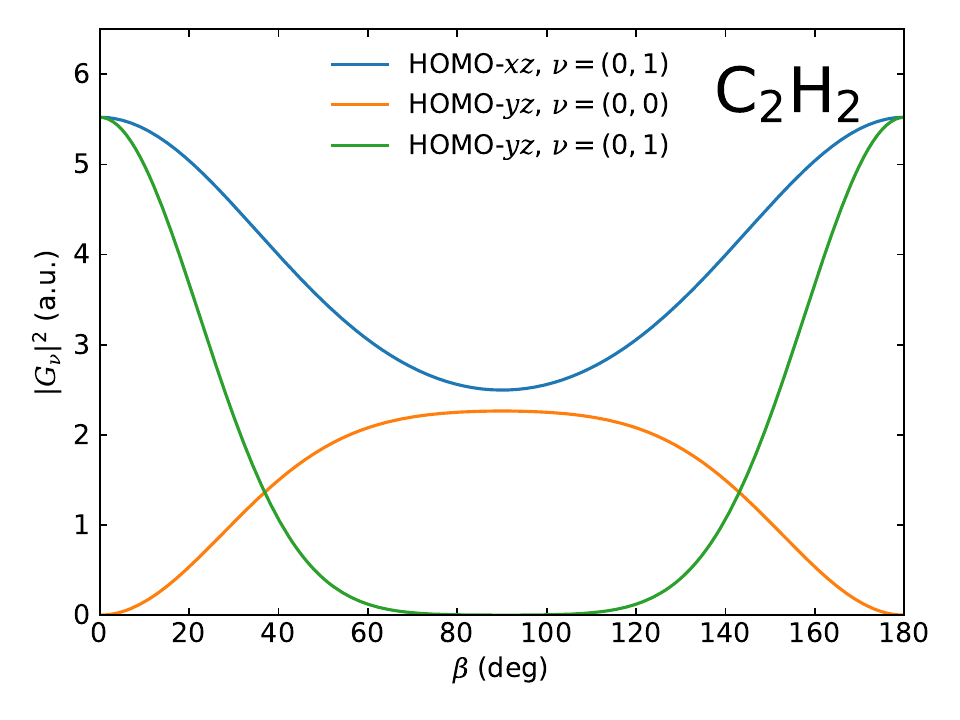}
    \caption{Squared structure factor $\abs{G_\nu}^2$ of the two degenerate HOMOs of the acetylene molecule ($\rm{C}_2 \rm{H}_2$). The molecular axis is aligned along the $z$-axis in the MF, the internuclear distance between the C nuclei is 1.20 \r{A}, and it is 1.06 \r{A} between the C nucleus and the neighboring H nucleus. The basis used for calculation is pc-2.}
    \label{fig:example_C2H2}
\end{figure}

\begin{figure}
    \centering
    \includegraphics[width=\columnwidth]{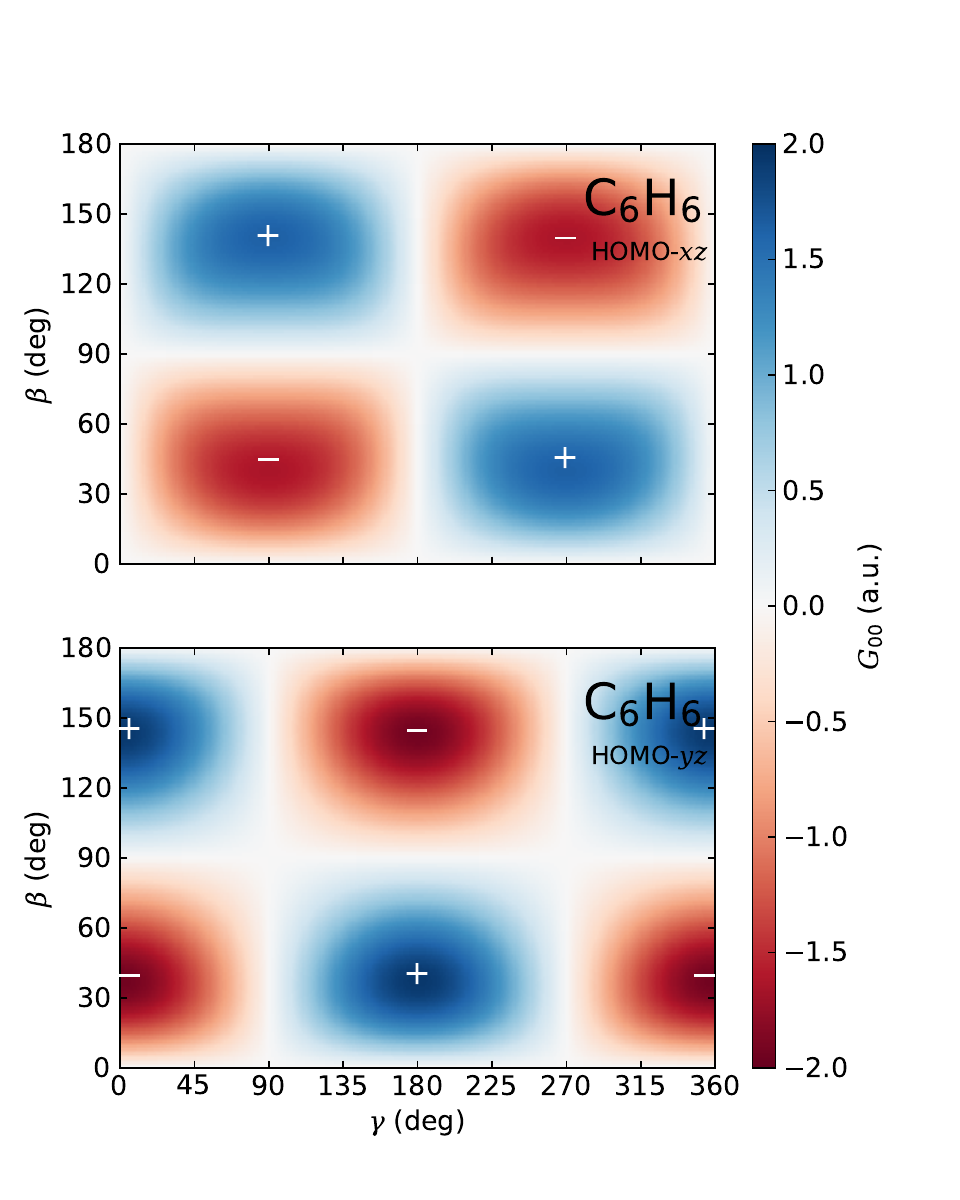}
    \caption{Structure factor $G_{00}$ of the two degenerate HOMOs of the benzene molecule ($\rm{C}_6 \rm{H}_6$). The molecule is placed in the $x$-$z$ plane in the MF, with the $z$-axis pointing from one C nucleus to its neighboring H nucleus. The basis used for calculation is pc-1.}
    \label{fig:example_C6H6}
\end{figure}

In this section we take a number of common molecules as examples, and present their structure factors obtained from \texttt{PyStructureFactor}. The examples cover molecules with various properties, as shown in Figs.~\ref{fig:example_H2} to \ref{fig:example_C6H6}: diatomic molecules ($\rm{H}_2$, $\rm{CO}$, and $\rm{O}_2$), linear polyatomic molecules ($\rm{C}_2\rm{H}_2$) and planar molecules ($\rm{C}_6\rm{H}_6$); open-shell molecules ($\rm{O}_2$); molecules with degenerate HOMOs (double HOMO degeneracy for $\rm{O}_2$, $\rm{C}_2\rm{H}_2$ and $\rm{C}_6\rm{H}_6$).

For all examples presented, we set the input parameter $l_{\rm{max}}$ to the default value of 10 and set the \texttt{``atom\_grid\_level''} parameter to the default value of 3.
For linear molecules, the $\gamma$ dependence of the structure factor drops out and it only depends on $\beta$. For nonlinear molecules, on the other hand, the structure factor is presented on the $(\beta,\gamma)$ grid, see Fig.~\ref{fig:example_C6H6} for an example of the $\rm{C}_6\rm{H}_6$ molecule.

It is worthwhile noting that the molecular structure factor varies not only from molecule to molecule, but also changes from orbital to orbital for a particular molecule. Novel laser technologies have enabled the exploration of ionization from different molecular orbitals, and studies show that subsequent molecular dynamics critically depends on the ionizing orbital. Not surprisingly, the structure factor also depends on the orbital from which ionization occurs, as shown in Figs.~\ref{fig:example_H2} and \ref{fig:example_CO}.

Furthermore, for molecules with degenerate orbitals, the structure factor depends on the orientation of the orbital. The HOMOs of $\rm{O}_2$, for example, are $\pi$ orbitals that are two degenerate orbitals perpendicular to each other. Each of them has a nodal plane, and we use the nodal plane to identify the two degenerate orbitals. ``HOMO-$yz$'' corresponds to the orbital that has a nodal plane on the $y$-$z$ plane while ``HOMO-$xz$'' represents that with a nodal plane on the $x$-$z$ plane. Clearly, the structure factor depends on the orientation of the orbital and thus varies for different degenerate HOMO orbitals, as shown in Figs.~\ref{fig:example_O2} and \ref{fig:example_C2H2}. We note that tunneling ionization could occur simultaneously from multiple orbitals, and the final tunneling ionization rate might be a summation over various contributions.

\section{Conclusions}
\label{sec:concl}

The tunneling ionization rate of molecules depends on the molecular orientation. Such dependence is determined by the molecular structure factor. In this work, we have presented the \texttt{PyStructureFactor} program, which computes the molecular structure factor within the WFAT theoretical framework based on the \texttt{PySCF} quantum chemistry package. \texttt{PyStructureFactor} is implemented in a manner that is both computationally efficient and easily accessible to the end users. It provides a straightforward approach to obtain the molecular structure factor of virtually all kinds of molecules, and therefore lays a solid foundation for the study of strong-field molecular dynamics.

\section*{Acknowledgments}

We would like to thank Qiming Sun, Hao Huang, and Shengzhe Pan for helpful discussions.
This work is supported by the National Natural Science Foundation of China (Grant Nos.\ 92150105, 11834004, 12227807, and 12241407) and the Science and Technology Commission of Shanghai Municipality (Grant No.\ 21ZR1420100).
Numerical computations were in part performed on the East China Normal University Multifunctional Platform for Innovation (001).

\bibliographystyle{elsarticle-num}
\bibliography{Article_PySF.bib}

\end{document}